\documentclass[11pt]{article}
\usepackage{amssymb}
\usepackage{amsthm}
\usepackage{amsmath}
\usepackage{setspace}
\usepackage{natbib}
\usepackage{pstricks}
\usepackage{pstricks-add}
\usepackage{xcolor}
\usepackage{subfig}
\usepackage{graphics}
\usepackage{multirow}
\usepackage{booktabs}
\usepackage{geometry}

\newenvironment{sistema}{\left\lbrace\begin{array}{@{}l@{}}}{\end{array}\right.}

\begin{document}

\title{Extending Yagil exchange ratio determination model to the case of stochastic dividends}

\author{Enrico Moretto\footnote{Dipartimento di Economia - Universit\`a dell'Insubria, via Monte Generoso 71, 21100 Varese, Italy, and CNR-IMATI, via A. Corti 12, 20133, Milano, Italy. E-mail address: enrico.moretto@uninsubria.it} \qquad \qquad Alessandra Mainini\footnote{Dipartimento di Discipline matematiche, Finanza matematica ed Econometria - Universit\`a Cattolica del Sacro Cuore - via L. Necchi 9, 20123, Milano, Italy. E-mail address: alessandra.mainini@unicatt.it}}
\date{}
\maketitle

\begin{abstract}
This article extends, in a stochastic environment, the \cite{Yagil} model which establishes, in a deterministic dividend discount model, a range for the exchange ratio in a \textit{stock-for-stock} merger agreement. Here, we generalize Yagil's work letting both pre- and post-merger dividends grow randomly over time. If Yagil focuses only on changes in stock prices before and after the merger, our stochastic environment allows to keep in account both shares' expected values and variance, letting us to identify a more complex bargaining region whose shape depends on mean and standard deviation of the dividends' growth rate.


\end{abstract}

\newpage

\section{Introduction, literature review and motivation}
\label{Intro}

Mergers and acquisitions have been, and still are, a widely studied topic in financial literature, under both a theoretical and an empirical point of view. Companies merge for various reasons, but with a unique goal: to create synergy, the additional equity value of the newly created company ($M$) when compared to the pre-existing ones, namely the acquiring ($A$) and the acquired, or target, ($B$). In a stock-for-stock merger, $B$'s shareholders receive, for each stock they give up, $r$ (the \textit{exchange ratio}) stocks of company $M$. Shareholders of companies $A$ and $B$ will agree on some value for $r$ only if their wealth increases after the merger. Negotiation on $r$  establishes the portion of synergy that goes to stockholders of pre-merger companies. It is therefore crucial to identify a \textit{bargaining region}, that is a non-empty range for $r$.

First attempts in this direction go back to \cite{LG} and \cite{Yagil}. Larson and Gonedes represent the value of all companies in terms of their price-earnings ratios, and determine the minimum and maximum $r$ acceptable for all shareholders in terms of $M$'s price-earnings ratio.

Yagil tackles the same issue using the dividend discount model (DDM) by \cite{Williams} and \cite{GS}. Here the price of a common stock is the sum of all discounted future dividends companies will pay to shareholders; Further, dividends are assumed to grow at a constant and deterministic rate. Yagil determines the bargaining region for each synergy generating $M$ dividends' growth rate.

In both these models, $A$ and $B$'s shareholders have conflicting interests: the acquiring (acquired) company aims at fixing $r$ as low (high) as possible.

\cite{MR} determine, in an equilibrium context, the exchange ratio in terms of the expected synergy created by the merger and the companies' riskiness, while \cite{TH} analyze the effects of a merger by means of utility theory.

This paper generalizes Yagil's model by exploiting the Stochastic Dividend Discount Model (SDDM) (\cite{HJ94}, \cite{HJ98}, \cite{Yao}, and  \cite{Hurley}). Future dividends are driven by a stochastic growth rate and evolve in a Markovian fashion. Along with an expression for the expected current stock price, recently, a formula for variance (\cite{AM}) and covariance between stock prices (\cite{AMM}) have been determined. A further step in this direction can be found in \cite{D'Amico2013}, \cite{D'Amico2016}, and \cite{BDD}, where stochastic dividends evolve according to a more general semi-Markov dynamics.

In our stochastic setting, shareholders accept to merge if they all benefit not only from an increase in the expected value of their random wealth but also from a reduction in its variance. The bargaining area, now a function of both mean and standard deviation of $M$ dividends' growth rate, shows that large values for this dividends' expected growth rate is not always good news as this quantity affects also company $M$ stock price variance. Stockholders might, consequentely, end up, after the merger, in a riskier position.

The paper is organized as follows.  Section \ref{SDDMYagil} describes the theoretical framework and determines the bargaining region in a SDDM setting, Section \ref{Examples} provides a numerical example,  Section \ref{Remarks} eventually concludes.

\section{A SDDM extension of Yagil's model}
\label{SDDMYagil}

The main assumption behind the stochastic extension of the Dividend Discount Model is that the total amount of dividends $\tilde D(t)$ a company pays in $t$ to its shareholders evolve through time by means of the stochastic recursive equation $\tilde D \left(t+1\right) = \tilde D \left(t\right) \left(1+\tilde g\right)$, being $D(0)$  the last paid certain dividend and $\tilde g$ the dividends' growth rate represented by the following finite-state random variable,
$$
\tilde g =
\begin{sistema}
\begin{array}{cccccc}
\mbox{rate of growth} & g_1 & g_2  & ...  & g_n  &  \\
\mbox{probability}    & p_1 & p_2  & ...  & p_n  &
\end{array}
\end{sistema}
$$
with $-1 < g_1 < ... < g_n$, $\mathbb P\left[\tilde g = g_s \right] > 0$, $s = 1, ..., n$, and $p_1 + ... + p_n =1$.

Subscript $i = A,B,M$ relates to the acquiring, acquired, and resulting  companies. We assume that each company is characterized by a specific distribution for $\tilde g$, with $\bar g_i$ and $\sigma_{\tilde g_i}$, respectively, its expected value and variance. Let $N_i$ denote the number of company $i$'s outstanding stocks and  $\tilde d_i(t) = \tilde D_i(t) / N_i$ its random dividends-per-share (\textit{dps}) at time $t$. The current random stock price is
\begin{equation}
\tilde P_i (0) = \sum_{t=1}^{+\infty} \frac{d_i(0) \left(1+\tilde g_i\right)^t}{\left(1+ k_i\right)^t},
\label{Prezzo}
\end{equation}
being $k_i$ company $i$ constant and deterministic risk-adjusted discount rate. Company $i$'s equity value is, then, $\tilde W_i(0) = \tilde P_i(0) N_i$.

$M$'s \textit{dps} in $0$ is
$$
d_M(0) = \frac{D_A(0) + D_B (0)}{N_A + r N_B}.
$$
and will grow according to $\tilde g_M$.

Hurley and Johnson (1994, 1998) and Yao (1997) prove that the expected stock price is, as long as $k_i > \bar g_i$,
\begin{equation}
\bar{P}_i(0) = \frac{d_i(0) \left(1 + \bar g_i\right)}{k_i - \bar g_i}. 
\end{equation}
Agosto and Moretto (2015) determine the stock price variance
\begin{equation*}
\sigma^2_i(0) =
\frac{\bar P_i^2(0)h\left(\bar g_i,\sigma_{\tilde g_i}\right)\left(1 + k_i\right)^2}{\left(1 + \bar g_i\right)^2},
\end{equation*}
being
\[
h\left(\bar g_i,\sigma_{\tilde g_i}\right) = \frac{\sigma_{\tilde g_i}}{\sqrt{\Delta_i }},
\hspace{0.35 cm} \sigma_{\tilde{g}_i} > 0,
\]
and where $\Delta_i = \left(1 + k_i\right)^2 - \left(1 + \bar g_i\right)^2 - \sigma^2_{\tilde{g}_i}$ has to be strictly positive. It will reveal handy to denote the coefficient of variation of $\tilde{P}_i\left(0\right)$ as
$$
f_i
= \frac{h\left(\bar g_i,\sigma_{\tilde g_i}\right)\left(1 + k_i\right)}{1 + \bar g_i}.
$$

The crucial assumption in Yagil is the choice of a deterministic growth rate for $M$. In his setting, the agreement is attainable if stockholders of both company $A$ and $B$ enjoy a positive gain in wealth, that is $P_M(0) \geq P_A(0)$ and $r P_M(0) \geq P_B(0)$,
being $P_i(0)$ the stock price of company $i$ resulting when a deterministic growth rate replaces $\tilde g_i$  in (\ref{Prezzo}).

Moreover, Yagil assumes that the discount rate of the resulting company is the weighted average of $k_A$ and $k_B$, with weights equal to the relative equity values. That is like saying that the merger does not influence the overall risk of the resulting company with respect of the pre-existing ones. Here, $k_M$ is calculated accordingly.

The SDDM generalization of Yagil's model assumes that shareholders of company $A$ (resp. $B$) are better off, in terms of expected values, when
\begin{equation}
\bar P_M(0) \geq \bar P_A(0) \; \; (\mbox{resp.} \; \; r \bar P_M(0) \geq \bar P_B(0)) \label{Cond1}
\end{equation}
and, in terms of variance, when
\begin{equation}
\sigma^2_M(0) \leq \sigma^2_A(0) \; \; (\mbox{resp.} \; \; r^2 \sigma^2_M(0) \leq \sigma^2_B(0)) \label{Cond2}
\end{equation}
hold. An increase in terms of expected wealth for both groups of shareholders (\textit{i.e.}, condition (\ref{Cond1}) holds) occurs when
\begin{equation}
\frac{N_A}{N_B}
	\frac{\bar W_B(0)}{\bar W_M(0) - \bar W_B(0)} \leq r \leq \frac{N_A}{N_B} \frac{\bar W_M(0) - \bar W_A(0)}{\bar W_A(0)},
\label{ExchangeRatio1}
\end{equation}
where $\bar W_i(0) = N_i\bar P_i(0)$. There is a reduction in  variance (\textit{i.e.}, condition (\ref{Cond2}) holds)  when
\begin{equation}
\frac{N_A}{N_B}
\frac{\bar W_M(0) f_M - \bar W_A(0) f_A}{\bar W_A(0) f_A}
 \leq r \leq
\frac{N_A}{N_B} \frac{\bar W_B(0) f_B}{\bar W_M(0) f_M - \bar W_B(0) f_B}.
\label{ExchangeRatio2}
\end{equation}
Interval (\ref{ExchangeRatio1}) is not empty when $\bar W_M(0) \geq \bar W_A(0) + \bar W_B(0)$ that is, the merger creates synergy with positive expected value; (\ref{ExchangeRatio1}) collapses to a unique point $r^{\ast} = \bar P_B(0) / \bar P_A(0)$ in case of no synergy, that is if $\bar W_M(0) = \bar W_A(0) + \bar W_B(0)$.

Interval (\ref{ExchangeRatio2}) is instead not empty when
\begin{equation}
   f_M\bar W_M \left(0\right)  \leq f_A \bar W_A \left(0\right) + f_B \bar  W_B \left(0\right). \label{CondVar}
\end{equation}
Condition (\ref{CondVar}) carries some interesting remarks. Firstly, as the coefficient of variation resembles the reciprocal of the Sharpe's ratio, shareholders should prefer stocks with smaller $f$, that is with larger risk premium (per unit of deviation). This means that if company $M$ guarantees a sufficiently large risk compensation, stockholders will benefit from a reduction in their wealth's variance. In case of no synergy, (\ref{CondVar}) becomes
\begin{equation}
f_M \leq \omega_Af_A + \omega_Bf_B,
\;\;\;\;\;
\omega_i = \frac{\bar W_i \left(0\right)}{\bar W_A \left(0\right) + \bar W_B \left(0\right)},\;\;i = A,B,
\label{MeanCV}
\end{equation}
whose rhs term is the weighted average of $f_A$ and $f_B$ with, as weights, the relative equity values of $A$ and $B$. Merger is, then, profitable if $M$ is less risky than an equity-valued `portfolio' of $A$ and $B$.

Unlike (\ref{ExchangeRatio1}), in case of no synergy interval (\ref{ExchangeRatio2}) does not collapse into a single value. Substituting $\bar W_M(0) = \bar W_A(0) + \bar W_B(0)$ into (\ref{ExchangeRatio2}) leads to
\begin{equation}
\frac{N_A}{N_B} \left(\frac{f_M}{f_A} - 1 \right) + \frac{\bar P_B(0)}{\bar P_A(0)} \frac{f_M}{f_A} \leq r \leq
\left(\frac{N_B}{N_A} \left(\frac{f_M}{f_B} - 1 \right) + \frac{\bar P_A(0)}{\bar P_B(0)} \frac{f_M}{f_B}\right)^{-1}.
\label{IntervalloSenzaSinergia}
\end{equation}
This interval shrinks to $r^{\ast}$ only when $f_M = f_A$ and $f_M = f_B$, the case in which $A$ and $B$ have the same Sharpe ratio and no risk reduction is possible.

Finally, it is easy to prove that the intersection between (\ref{ExchangeRatio1}) and (\ref{ExchangeRatio2}) is not empty if $f_M \leq \min\left(f_A;f_B\right)$; that is, the new company is even less risky than the less risky of both $A$ and $B$, a situation that guarantees proper diversification. This condition  also ensures that (\ref{MeanCV}) holds so that (\ref{IntervalloSenzaSinergia}) contains, at least, $r^{\ast}$.

\section{A numerical example}
\label{Examples}

To better understand the effects of SDDM on the pre-merger negotiation, thus highlighting the difference with Yagil's setting, we consider a numerical example where the combined effect of $\bar g_M$ and $\sigma_{\tilde g_M }$ is studied. This allows to check if a negotiation is possible, and how easily the two parties will conclude a merging agreement. We assume that the larger the region defined simultaneously by (\ref{ExchangeRatio1}) and (\ref{ExchangeRatio2}) the `simpler' the agreement will be.

In our general setting, the extrema of intervals (\ref{ExchangeRatio1}) and (\ref{ExchangeRatio2}) are monotonic with respect to $\bar g_M$ and $\sigma_{\tilde g_M}$. If we define the constant
\[
H_i = \frac{D_A + D_B}{\bar W_i(0)} \geq0 ,\;\;\;i = A,B,
\]
interval (\ref{ExchangeRatio1}) can be rewritten as
\[
\frac{N_A}{N_B}\left(\frac{1 + \bar g_M}{k_M - \bar g_M}H_B - 1\right)^{-1}
\leq r \leq
\frac{N_A}{N_B}\left(\frac{1 + \bar g_M}{k_M - \bar g_M}H_A - 1\right).
\]
The infimum (resp. the supremum) of this interval decreases (resp. increases) in $\bar g_M$; the bargaining region defined by (\ref{Cond1}) becomes larger because the expected wealth of  shareholders of both companies increases; concluding an agreement becomes easier. If we, instead, define the constant
\[
J_i = \frac{(1 + k_M)H_i}{f_i}\geq0,\;\;\;i = A,B,
\]
the region defined by (\ref{Cond2}) can be written as
\[
\frac{N_A}{N_B}\left(\frac{h\left(\bar g_M,\sigma_{\tilde g_M}\right)}{k_M - \bar g_M}J_A  - 1\right)
\leq r \leq
\frac{N_A}{N_B}\left(\frac{h\left(\bar g_M,\sigma_{\tilde g_M}\right)}{k_M - \bar g_M}J_B  - 1\right)^{-1}.
\]
Again, it is straightforward to prove that the infimum (resp. the supremum) of this interval increases (resp. decreases) both in $\bar g_M$ (for each positive $\sigma_{\tilde g_M}$) and $\sigma_{\tilde g_M}$ (for each $\bar g_M > - 1$). Here, room for negotiation diminishes if the post-merger standard deviation $\sigma_{\tilde g_M}$ increases because it becomes difficult to achieve a lower post-merger risk. An increase in $\bar g_M$ has the same effect; this is so because the mean is the value that minimizes the centered second order moment. Quite interestingly, and somehow counter-intuitively, a variation in $\bar g_M$ has two opposite consequences on the region of negotiation, the overall result depending on which effect is dominating.

\begin{table}
\centering
\subfloat[Parameters]
{
\scalebox{0.9}{
\begin{tabular}{cccccc}
\toprule
$i$ & $d_i(0)$ & $\bar g_i$ & $\sigma_{\tilde g_i}$ & $k_i$ & $N_i$\\
\midrule
$A$ & $0.6$ & $1\%$ & $2\%$ & $4\%$ & $1\ 000$ \\
\midrule
$B$ & $0.3$ & $3\%$ & $9\%$ & $8\%$ & $2\ 500$ \\
\bottomrule
\end{tabular}}}
\quad
\subfloat[SDDM results]
{
\scalebox{0.9}{
\begin{tabular}{cccccc}
\toprule
$i$ & $\bar P_i$ & $\sigma_i$ & $f_i$ & $\bar W_i$ & $\omega_i$ \\
\midrule
$A$ & $20.2$ & $1.68$ & $0.0832$ & $20\ 200$ &  $0.57$ \\
\midrule
$B$ & $6.18$ & $1.87$ & $0.3026$ & $15\ 450$ & $0.43$ \\
\bottomrule
\end{tabular}}}
\caption{Companies $A$ and $B$ pre-merging values}
\label{tabelle}
\end{table}

Table (\ref{tabelle}.a) depicts the parameters describing pre-merger companies $A$ and $B$ while Table (\ref{tabelle}.b) reports their SDDM relevant values (stock prices mean and standard deviation, coefficient of variation, absolute ($\bar W_i$) and relative ($\omega_i$) equity values). As $f_B > f_A$, the target company is riskier than the acquiring. According to Yagil, the discount rate for $M$ is $5.72\%$.

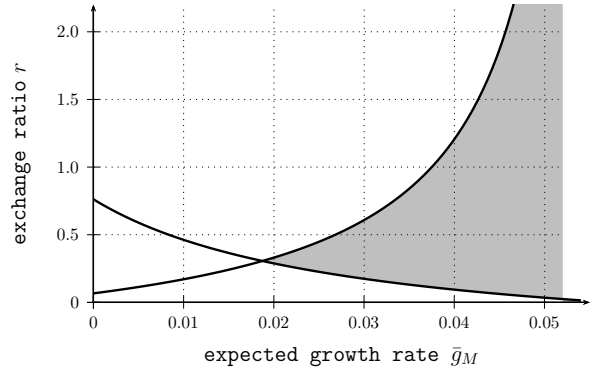
\begin{figure}
\centering 
\psset{plotpoints=100,algebraic,xunit=200,yunit=3}
\scalebox{0.6}{
\begin{pspicture*}
(-0.02,-0.5)(.062,2.2)
\rput(.0275,-0.4){\Large{\texttt{expected growth rate $\bar g_M$}}}
\rput(-0.008,1.1){\Large{\rotateleft{{\texttt{exchange ratio} $r$}}}}
\pscustom[fillstyle=solid,fillcolor=lightgray,linestyle=none]{
\psplot{0.0188}{.054}{0.4*(((13.5/154.5)*(1 + x)/(0.0573 - x) - 1)^(-1))},
\psplot{.054}{0.052}{0},
\psline(0.052,0)(0.052,2.8),
\psplot{0.052}{.049}{2.8},
\psplot{.049}{0.0188}{0.4*((13.5/202)*(1 + x)/(0.0573 - x) - 1)}}
\psplot[linewidth=1.5pt]{0}{.054}{0.4*((13.5/202)*(1 + x)/(0.0573 - x) - 1)}
\psplot[linewidth=1.5pt]{0}{.054}{0.4*(((13.5/154.5)*(1 + x)/(0.0573 - x) - 1)^(-1))}
\psaxes[dx =0.01,dy = 0.5,Dx=0.01,Dy=0.5]{->}(0,0)(0,0)(.055,2.2)
\psline[linestyle=dotted](0,0.5)(.052,0.5)
\psline[linestyle=dotted](0,1)(.052,1)
\psline[linestyle=dotted](0,1.5)(.052,1.5)
\psline[linestyle=dotted](0,2)(.052,2)
\psline[linestyle=dotted](0,2.5)(.052,2.5)
\psline[linestyle=dotted](0.01,0)(0.01,2.2)
\psline[linestyle=dotted](0.02,0)(0.02,2.2)
\psline[linestyle=dotted](0.03,0)(0.03,2.2)
\psline[linestyle=dotted](0.04,0)(0.04,2.2)
\psline[linestyle=dotted](0.05,0)(0.05,2.2)
\end{pspicture*}}
\caption{The case studied in Yagil (1987).}
\label{bargaining-region-Yagil}
\end{figure}

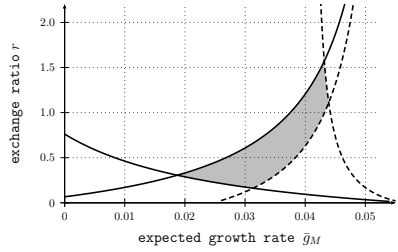
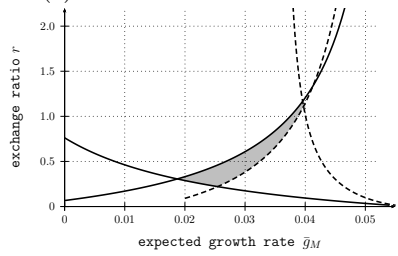
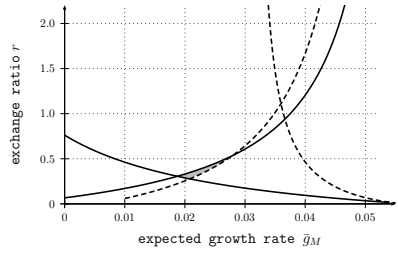
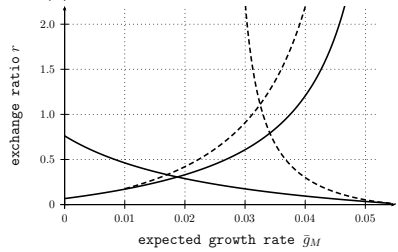
\begin{figure}
\centering
\subfloat[][Standard deviation: $1\%$]
{\psset{plotpoints=100,algebraic,xunit=200,yunit=3}
\scalebox{0.4}{
\begin{pspicture*}
(-0.02,-0.5)(.062,2.2)
\psaxes[dx =0.01,dy = 0.5,Dx=0.01,Dy=0.5]{->}(0,0)(0,0)(.055,2.2)
\rput(.0275,-0.4){\Large{\texttt{expected growth rate $\bar g_M$}}}
\rput(-0.008,1.1){\Large{\rotateleft{{\texttt{exchange ratio} $r$}}}}
\pscustom[fillstyle=solid,fillcolor=lightgray,linestyle=none]{
\psplot{0.0188}{.031}{0.4*(((13.5/154.5)*(1 + x)/(0.0573 - x) - 1)^(-1))},
\psplot{0.031}{.044}{0.4*((13.5/202)*(1.0573/0.0833)*(1/(0.0573 - x))*(((0.01^2)/((1.0573^2) - ((1 + x)^2) - (0.01^2)))^(0.5)) - 1)},
\psplot{0.044}{.043}{0.4*(((13.5/154.5)*(1.0573/0.3024)*(1/(0.0573 - x))*(((0.01^2)/((1.0573^2) - ((1 + x)^2) - (0.01^2)))^(0.5)) - 1)^(-1))},
\psplot{.043}{0.0188}{0.4*((13.5/202)*(1 + x)/(0.0573 - x) - 1)}}
\psplot[linewidth=1.5pt]{0}{.048}{0.4*((13.5/202)*(1 + x)/(0.0573 - x) - 1)}
\psplot[linewidth=1.5pt]{0}{.054}{0.4*(((13.5/154.5)*(1 + x)/(0.0573 - x) - 1)^(-1))}
\psplot[linewidth=1.5pt,linestyle=dashed]{0.026}{.054}{0.4*((13.5/202)*(1.0573/0.0833)*(1/(0.0573 - x))*(((0.01^2)/((1.0573^2) - ((1 + x)^2) - (0.01^2)))^(0.5)) - 1)}
\psplot[linewidth=1.5pt,linestyle=dashed]{0.042}{.055}{0.4*(((13.5/154.5)*(1.0573/0.3024)*(1/(0.0573 - x))*(((0.01^2)/((1.0573^2) - ((1 + x)^2) - (0.01^2)))^(0.5)) - 1)^(-1))}
\psline[linestyle=dotted](0,0.5)(.052,0.5)
\psline[linestyle=dotted](0,1)(.052,1)
\psline[linestyle=dotted](0,1.5)(.052,1.5)
\psline[linestyle=dotted](0,2)(.052,2)
\psline[linestyle=dotted](0,2.5)(.052,2.5)
\psline[linestyle=dotted](0.01,0)(0.01,2.2)
\psline[linestyle=dotted](0.02,0)(0.02,2.2)
\psline[linestyle=dotted](0.03,0)(0.03,2.2)
\psline[linestyle=dotted](0.04,0)(0.04,2.2)
\psline[linestyle=dotted](0.05,0)(0.05,2.2)
\end{pspicture*}}}
\quad
\subfloat[][Standard deviation: $1.5\%$]
{\psset{plotpoints=100,algebraic,xunit=200,yunit=3}
\scalebox{0.4}{
\begin{pspicture*}
(-0.02,-0.5)(.062,2.2)
\psaxes[dx =0.01,dy = 0.5,Dx=0.01,Dy=0.5]{->}(0,0)(0,0)(.055,2.2)
\rput(.0275,-0.4){\Large{\texttt{expected growth rate $\bar g_M$}}}
\rput(-0.008,1.1){\Large{\rotateleft{{\texttt{exchange ratio} $r$}}}}
\pscustom[fillstyle=solid,fillcolor=lightgray,linestyle=none]{
\psplot{0.0188}{.025}{0.4*(((13.5/154.5)*(1 + x)/(0.0573 - x) - 1)^(-1))},
\psplot{0.025}{.04}{0.4*((13.5/202)*(1.0573/0.0833)*(1/(0.0573 - x))*(((0.015^2)/((1.0573^2) - ((1 + x)^2) - (0.015^2)))^(0.5)) - 1)},
\psplot{0.04}{.0395}{0.4*(((13.5/154.5)*(1.0573/0.3024)*(1/(0.0573 - x))*(((0.015^2)/((1.0573^2) - ((1 + x)^2) - (0.015^2)))^(0.5)) - 1)^(-1))},
\psplot{.0395}{0.0188}{0.4*((13.5/202)*(1 + x)/(0.0573 - x) - 1)}}
\psplot[linewidth=1.5pt]{0}{.048}{0.4*((13.5/202)*(1 + x)/(0.0573 - x) - 1)}
\psplot[linewidth=1.5pt]{0}{.054}{0.4*(((13.5/154.5)*(1 + x)/(0.0573 - x) - 1)^(-1))}
\psplot[linewidth=1.5pt,linestyle=dashed]{0.02}{.055}{0.4*((13.5/202)*(1.0573/0.0833)*(1/(0.0573 - x))*(((0.015^2)/((1.0573^2) - ((1 + x)^2) - (0.015^2)))^(0.5)) - 1)}
\psplot[linewidth=1.5pt,linestyle=dashed]{0.037}{.055}{0.4*(((13.5/154.5)*(1.0573/0.3024)*(1/(0.0573 - x))*(((0.015^2)/((1.0573^2) - ((1 + x)^2) - (0.015^2)))^(0.5)) - 1)^(-1))}
\psline[linestyle=dotted](0,0.5)(.052,0.5)
\psline[linestyle=dotted](0,1)(.052,1)
\psline[linestyle=dotted](0,1.5)(.052,1.5)
\psline[linestyle=dotted](0,2)(.052,2)
\psline[linestyle=dotted](0,2.5)(.052,2.5)
\psline[linestyle=dotted](0.01,0)(0.01,2.2)
\psline[linestyle=dotted](0.02,0)(0.02,2.2)
\psline[linestyle=dotted](0.03,0)(0.03,2.2)
\psline[linestyle=dotted](0.04,0)(0.04,2.2)
\psline[linestyle=dotted](0.05,0)(0.05,2.2)
\end{pspicture*}}} \\
\subfloat[][Standard deviation: $2\%$]
{\psset{plotpoints=100,algebraic,xunit=200,yunit=3}
\scalebox{0.4}{
\begin{pspicture*}
(-0.02,-0.5)(.062,2.2)
\psaxes[dx =0.01,dy = 0.5,Dx=0.01,Dy=0.5]{->}(0,0)(0,0)(.055,2.2)
\rput(.0275,-0.4){\Large{\texttt{expected growth rate $\bar g_M$}}}
\rput(-0.008,1.1){\Large{\rotateleft{{\texttt{exchange ratio} $r$}}}}
\pscustom[fillstyle=solid,fillcolor=lightgray,linestyle=none]{
\psplot{0.0188}{.02}{0.4*(((13.5/154.5)*(1 + x)/(0.0573 - x) - 1)^(-1))},
\psplot{0.02}{.027}{0.4*((13.5/202)*(1.0573/0.0833)*(1/(0.0573 - x))*(((0.02^2)/((1.0573^2) - ((1 + x)^2) - (0.02^2)))^(0.5)) - 1)},
\psplot{.027}{0.0188}{0.4*((13.5/202)*(1 + x)/(0.0573 - x) - 1)}}
\psplot[linewidth=1.5pt]{0}{.048}{0.4*((13.5/202)*(1 + x)/(0.0573 - x) - 1)}
\psplot[linewidth=1.5pt]{0}{.054}{0.4*(((13.5/154.5)*(1 + x)/(0.0573 - x) - 1)^(-1))}
\psplot[linewidth=1.5pt,linestyle=dashed]{0.01}{.045}{0.4*((13.5/202)*(1.0573/0.0833)*(1/(0.0573 - x))*(((0.02^2)/((1.0573^2) - ((1 + x)^2) - (0.02^2)))^(0.5)) - 1)}
\psplot[linewidth=1.5pt,linestyle=dashed]{0.033}{.055}{0.4*(((13.5/154.5)*(1.0573/0.3024)*(1/(0.0573 - x))*(((0.02^2)/((1.0573^2) - ((1 + x)^2) - (0.02^2)))^(0.5)) - 1)^(-1))}
\psline[linestyle=dotted](0,0.5)(.052,0.5)
\psline[linestyle=dotted](0,1)(.052,1)
\psline[linestyle=dotted](0,1.5)(.052,1.5)
\psline[linestyle=dotted](0,2)(.052,2)
\psline[linestyle=dotted](0,2.5)(.052,2.5)
\psline[linestyle=dotted](0.01,0)(0.01,2.2)
\psline[linestyle=dotted](0.02,0)(0.02,2.2)
\psline[linestyle=dotted](0.03,0)(0.03,2.2)
\psline[linestyle=dotted](0.04,0)(0.04,2.2)
\psline[linestyle=dotted](0.05,0)(0.05,2.2)
\end{pspicture*}}}
\quad
\subfloat[][Standard deviation: $2.5\%$]
{\psset{plotpoints=100,algebraic,xunit=200,yunit=3}
\scalebox{0.4}{
\begin{pspicture*}
(-0.02,-0.5)(.062,2.2)
\psaxes[dx =0.01,dy = 0.5,Dx=0.01,Dy=0.5]{->}(0,0)(0,0)(.055,2.2)
\rput(.0275,-0.4){\Large{\texttt{expected growth rate $\bar g_M$}}}
\rput(-0.008,1.1){\Large{\rotateleft{{\texttt{exchange ratio} $r$}}}}
\psplot[linewidth=1.5pt]{0}{.048}{0.4*((13.5/202)*(1 + x)/(0.0573 - x) - 1)}
\psplot[linewidth=1.5pt]{0}{.054}{0.4*(((13.5/154.5)*(1 + x)/(0.0573 - x) - 1)^(-1))}
\psplot[linewidth=1.5pt,linestyle=dashed]{0.01}{.045}{0.4*((13.5/202)*(1.0573/0.0833)*(1/(0.0573 - x))*(((0.025^2)/((1.0573^2) - ((1 + x)^2) - (0.025^2)))^(0.5)) - 1)}
\psplot[linewidth=1.5pt,linestyle=dashed]{0.028}{.055}{0.4*(((13.5/154.5)*(1.0573/0.3024)*(1/(0.0573 - x))*(((0.025^2)/((1.0573^2) - ((1 + x)^2) - (0.025^2)))^(0.5)) - 1)^(-1))}
\psline[linestyle=dotted](0,0.5)(.052,0.5)
\psline[linestyle=dotted](0,1)(.052,1)
\psline[linestyle=dotted](0,1.5)(.052,1.5)
\psline[linestyle=dotted](0,2)(.052,2)
\psline[linestyle=dotted](0,2.5)(.052,2.5)
\psline[linestyle=dotted](0.01,0)(0.01,2.2)
\psline[linestyle=dotted](0.02,0)(0.02,2.2)
\psline[linestyle=dotted](0.03,0)(0.03,2.2)
\psline[linestyle=dotted](0.04,0)(0.04,2.2)
\psline[linestyle=dotted](0.05,0)(0.05,2.2)
\end{pspicture*}}}
\caption{The effect of the growth rate's standard deviation on the bargaining region.}
\label{bargaining-region}
\end{figure}

As a benchmark, if $g_M$ replaces $\bar g_M$ Figure \ref{bargaining-region-Yagil} presents, in the plane $(\bar g_M,r)$, the Yagil's bargaining region, defined by the extrema of interval (\ref{ExchangeRatio1}); each point belonging to the region between the two curves, on the right of their intersection point, is such that the stock price of the new company satisfy shareholders of both companies, being admissible for the negotiation.  In Figure \ref{bargaining-region} we fix four levels of $\sigma_{\tilde g_M}$, namely $1\%$, $1.5\%$, $2\%$, $2.5\%$, and superimpose, for each of them, extrema of interval (\ref{ExchangeRatio2}) (dashed curves) on the solid curves of Figure \ref{bargaining-region-Yagil}, which still depict extrema of interval (\ref{ExchangeRatio1}). Each point of the region bounded by the two dashed curves on the left of their intersection point fulfills shareholders' requirements of a smaller wealth variance. The shaded area in each of the four plots in Figure \ref{bargaining-region} represents the overall resulting  bargaining region, which can eventually be empty (Figure \ref{bargaining-region}.d). Looking at each plot, it results evident that an increase in $\sigma_{\tilde g_M}$  reduces the possibility of negotiation and positively concluding an agreement becomes increasingly difficult. Indeed, the example shows how negotiation does not even take place with $\sigma_{\tilde g_M} = 2.5\%$ as $A$'s shareholders will not accept an excessive post-merger increase in their wealth variance.

All regions represented in Figure \ref{bargaining-region} shows that  negotiation can take place only if $\bar g_M \geq 1.88\%$. Further, if $\bar g_M = 1.88\%$, that is the merger creates no synergy, then the unique acceptable exchange ratio is $r^{\ast} = 0.3059$. This level is larger than the pre-merger company $A$'s expected rate, $\bar g_A = 1\%$ (Table \ref{tabelle}.a). Therefore, as long as $\sigma_{\tilde g_M}$ is sufficiently small stockholders of the acquiring company can accept exchange ratios larger than $1$ (Figures \ref{bargaining-region}.a and  \ref{bargaining-region}.b). On the other hand, $\bar g_M$ can be way smaller than $\bar g_B$ as the merger will reward $B$'s stockholders with a sharp reduction in the  standard deviation of their wealth. In fact, the numerical examples shows that negotiation takes place when $\sigma_{\tilde g_M}$ is far smaller then $\sigma_{\tilde g_B} = 9\%$. $B$'s shareholders accept small exchange ratios; in Figures \ref{bargaining-region}.a, \ref{bargaining-region}.b, and \ref{bargaining-region}.c the minimum accepted rate is always less than $0.5$ because the reduction in the expected dividends' growth rate is adequately rewarded with a smaller level of risk. Lastly, as long as $\sigma_{\tilde g_M}$ increases, the minimum $r$ accepted by $B$ increases whereas the maximum $r$ offered by $A$ decreases. This occurs until the recuction in standard deviation is no more sufficient to satisfy shareholders' requests.

\section{Concluding remarks}
\label{Remarks}

This article deals with exchange ratio determination model by Yagil and tries to extend it into a stochastic framework where both expected value and variance of stockholders' wealth have to be considered when evaluating a plausible range for the exchange ratio in stock-for-stock merger agreements. It turns out that dividends' rate of growth of the company that the merger creates plays a double, conflicting role. In fact, such growth rate is responsible for changes in both the expected value and variance of stockholders' wealth. It is not always true, at least in this framework, that merging companies should uniquely strive for a large post-merger growth rate as an augmented wealth  variance might suggest to either shareholders of the acquired or acquiring companies, or possibly to both groups, not to accept the agreement.


\end{document}